\documentclass{article}
\usepackage{logic-article,amsmath,enumerate,xspace,graphicx}
\newcommand{\wmsou}{{\sc wmso+u}\xspace}
\newcommand{\msou}{{\sc mso+u}\xspace}
\newcommand{\mso}{{\sc mso}\xspace}
\newcommand{\tree}[1]{\mathsf{tree}(#1)}
\begin{document}
	
	\title{The \msou theory of $(\Nat,<)$ is undecidable}
	\author{Miko{\l}aj Boja\'nczyk, Pawe{\l} Parys, Szymon Toru\'nczyk}
	\maketitle
	\begin{abstract}
		We consider the logic \msou, which is monadic second-order logic extended with the unbounding quantifier. The unbounding quantifier is used to say that a property of finite sets holds for sets of arbitrarily large size. We prove that the logic is undecidable on infinite words, i.e.~the \msou theory of $(\Nat,\le)$ is undecidable. This settles an open problem about the logic, and improves a previous undecidability result, which used infinite trees and additional axioms from set theory.
	\end{abstract}

\newcommand{\vectorh}{\mathbf h}
\newcommand{\vectorg}{\mathbf g}
\newcommand{\vectorf}{\mathbf f}
\newcommand{\vectorid}{\mathbf{id}}

\section{Introduction}
A celebrated result of B\"uchi is that the monadic second-order \mso theory is decidable for the structure of natural numbers with order
\begin{align*}
	(\Nat,\le).
\end{align*}
In other words, \mso is decidable for infinite words. This paper shows that the decidability fails after \mso is extended  with the unbounding quantifier. The unbounding quantifier, denoted by
\begin{align*}
\mathsf{U} X.\  \varphi(X),
\end{align*}
says that $\varphi(X)$ holds for arbitrarily large finite sets $X$. As usual with quantifiers, the formula $\varphi(X)$ might have other free variables except for $X$. Call \msou the extension of \mso by this quantifier. The   main contribution of the paper is  the following theorem.

\begin{theorem}\label{thm:main}
	The \msou theory of $(\Nat,\le)$ is undecidable. 
\end{theorem}

\paragraph*{Background.}  The logic \msou was  introduced in~\cite{DBLP:conf/csl/Bojanczyk04}, where it was shown that satisfiability is decidable for formulas on infinite trees where the $\mathsf U$~quantifier is used once and not under the scope of set quantification. The decidability result from~\cite{DBLP:conf/csl/Bojanczyk04} straightforwardly entails  decidability of the finite model problem for modal $\mu$-calculus with backward modalities. A significantly more powerful fragment of the logic, albeit for infinite words, was shown decidable in~\cite{DBLP:conf/lics/BojanczykC06} using automata with counters.  These automata  where further developed into the theory of cost functions initiated by Colcombet in~\cite{DBLP:conf/icalp/Colcombet09}. The decidability result  from~\cite{DBLP:conf/lics/BojanczykC06}  straightforwardly entails decidability of  the star height problem.

The difficulty of \msou comes from the interaction between the unbounding quantifier and quantification over possibly infinite sets. This motivated the study of \wmsou, which is the variant of \msou where set quantification is restricted to finite set. On infinite words, satisfiability of \wmsou is decidable, and the logic has an automaton model~\cite{DBLP:journals/mst/Bojanczyk11}. Similar results hold for infinite trees~\cite{DBLP:conf/stacs/BojanczykT12}. The results from~\cite{DBLP:conf/stacs/BojanczykT12} have been used to decide properties of {\sc ctl*}~\cite{DBLP:conf/concur/CarapelleKL13}. Currently, the strongest decidability result in this line  is about \wmsou on infinite trees extended with quantification over infinite paths~\cite{DBLP:conf/icalp/Bojanczyk14a}. The latter result entails decidability of problems such as the realisability problem for prompt {\sc ltl}~\cite{KupfermanPitermanVardi09}, deciding the winner in cost parity games~\cite{DBLP:conf/fsttcs/FijalkowZ12}, or deciding certain properties of energy games~\cite{DBLP:conf/cav/BrazdilCKN12}.

While the above results showed that fragments \msou can be decidable, and can be used to prove results not directly related to the logic, it was not known if the full logic was decidable.  The first evidence that \msou can be too expressive was given in~\cite{DBLP:journals/fuin/HummelS12}, where it was shown that \msou can define languages of infinite words that are arbitrarily high in the projective hierarchy from descriptive set~theory. This result was used in~\cite{DBLP:conf/icalp/BojanczykGMS14}, where it was shown that, modulo a certain assumption from set theory (namely~{\sc v=l}), the \msou theory of the complete binary tree is undecidable. The result from~\cite{DBLP:conf/icalp/BojanczykGMS14} implies that there can be no algorithm which decides \msou on the complete binary tree, and which has a correctness proof in the {\sc zfc} axioms of set theory. This paper strengthens the result from~\cite{DBLP:conf/icalp/BojanczykGMS14} in two ways: first, we use no additional assumptions from set theory, and second, we prove undecidability for words and not trees.

\section{Vector sequences}
Define a \emph{number sequence} to be an element of $\Nat^\omega$, and define a \emph{vector sequence}  to be an element of $(\Nat^*)^\omega$, i.e.~an infinite sequence of vectors of natural numbers of possibly different dimensions.   We write $\vectorf,\vectorg$ for vector sequences and $f,g$ for number sequences.  If $f$ is a number sequence and $\vectorf$ is a vector sequence, then we write  $f \in \vectorf$ if for every position $i$, the $i$-th number in the sequence $f$ appears in one of the coordinates of the $i$-th vector in the vector sequence $\vectorf$. 
Number sequences are called \emph{asymptotically equivalent} if they are bounded on the same sets of positions. A vector sequence $\vectorf$ is called an  \emph{asymptotic mix} of a vector sequence $\vectorg$ if every $f \in \vectorf$ is asymptotically equivalent to some $g \in \vectorg$.   A vector sequence of dimension $d$ is one where all   vectors have dimension $d$.

	In the proofs below we use the following definition:  two vector sequences are asymptotically equivalent if they have the same dimension $d$, and for each coordinate $i \in \set{1,\ldots,d}$ the corresponding number sequences are asymptotically equivalent.

	\begin{lemma}\label{lem:no-mix}
		Let $d \in \Nat$.  There exists a vector sequence of dimension $d$ which is not an asymptotic mix of any vector sequence of dimension $d-1$. 
	\end{lemma}
\begin{pr}
	The definition of asymptotic mix does not use the order structure of natural numbers, and therefore in the proof of  this lemma we allow sequences to be indexed by  other countable sets, namely vectors of natural numbers. By induction on $d$, we will prove the following claim about vector sequences indexed by $\Nat^d$. We claim that the $d$-dimensional  identity
	\begin{align*}
		\vectorid : \Nat^d \to \Nat^d,
	\end{align*}
	is not an assymptotic mix of any vector sequence
	\begin{align*}
		\vectorg : \Nat^d \to \Nat^{d-1}.
	\end{align*}
	The induction base of $d=1$ is vacuous. Let us prove the claim for dimension $d$ assuming that it has been proved for smaller dimensions. 
	
	Toward a contradiction, suppose that  the $d$-dimensional identity is an asymptotic mix of some $\vectorg : \Nat^{d} \to \Nat^{d-1}$. 	
	 Consider the  subset of arguments $\{0\} \times \Nat^{d-1}$. The first coordinate of the $d$-dimensional identity is bounded  on this subset, namely it is zero,  and therefore there  must be some $ g \in \vectorg$ which is bounded on this set. 	  Without loss of generality, we  assume that the first coordinate of $\vectorg$ is bounded on  arguments from $\{0\} \times \Nat^{d-1}$. Let
	 \begin{align*}
	 	\vectorg' : \Nat^d \to \Nat^{d-2}
	 \end{align*}
	 be the vector sequence obtained from $\vectorg$ by removing the first coordinate.
	 Let
	 \begin{align*}
	 	\pi_i : \Nat^d \to \Nat \qquad \mbox{with $i \in \set{2,\ldots,d}$}
	 \end{align*}
	 be the projection onto the $i$-th coordinate, which satisfies $\pi_i \in \vectorid$. Therefore, each $\pi_i$ must be asymptotically equivalent to some  $g_i \in \vectorg$. 	  Let $X_i \subseteq \Nat^d$ be the set of arguments $x$ where $g_i$ agrees with the first coordinate of $\vectorg$. In other words, when restricted to arguments outside $X_i$ the projection $\pi_i$ is asymptotically equivalent to some  $g_i \in \vectorg'$. Since the first coordinate of $\vectorg$ is bounded on the set $\set{0} \times X^{d-1}$, it follows that there is some $c_i \in \Nat$ such that $X_i$ does not contain any arguments which have zero on the  first coordinate zero and at least $c_i$ on the  $i$-th coordinate.  Taking $c$ to be the maximum of all $c_2,\ldots,c_d$, we see that none of the sets $X_2,\ldots,X_d$ intersect the set
	 \begin{align*}
	 	X = \set{(0,n_2,\ldots,n_d) : n_2,\ldots,n_d \ge c}.
	 \end{align*}
It is easy to observe that the vector sequence
\begin{align}\label{eq:smaller-dim}
	(0,n_2,\ldots,n_d) \in X \qquad \mapsto \qquad (n_2,\ldots,n_d)
\end{align}
is an asymptotic mix of  $\vectorg'$, which is a vector sequence of dimension $d-2$. This contradicts the induction assumption, because
 the vector sequence in~\eqref{eq:smaller-dim} is asymptotically equivalent to the $(d-1)$-dimensional identity.
\end{pr}

A vector sequence is said to have \emph{bounded dimension} if there is some $d$ such that all vectors in the sequence have dimension at most $d$. A vector sequence is said to \emph{tend to infinity} if for every $n$, all but finitely many vectors in the sequence have all coordinates at least $n$.
We order vector sequences  coordintewise in the following way: we write $\vectorf \le \vectorg$ if for every $i$, the $i$-th vectors in both sequences have the same dimension, and the $i$-th vector of $\vectorf$  is coordinstewise smaller or equal to the $i$-th vector of $\vectorg$. A corollary of the above lemma is the following lemma, which characterises dimensions in terms only of boundedness properties.

\begin{lemma}\label{lem:dimension-char}
	Let $\vectorf_1,\vectorf_2$ be vector sequences of bounded dimensions which tend to infinity. Then following conditions are equivalent
	\begin{enumerate}
		\item on infinitely many positions $\vectorf_1$ has a vector of higher dimension than $\vectorf_2$;
		\item there exists some $\vectorg_1 \le \vectorf_1$ which is not an asymptotic mix of any $\vectorg_2 \le \vectorf_2$.
	\end{enumerate}  
\end{lemma}
\begin{pr}	
	Vector sequences that tend to infinity are maximal with respect to asymptotical equivalence in the following sense:  if a vector sequence $\vectorf$  of fixed dimension $d$ tends to infinity, then for every vector sequence $\vectorh$ of same dimension there exists an asymptotically equivalent vector sequence $\vectorg \le \vectorf$
	(to obtain such $\vectorg$, on each coordinate of each position we can take the minimum of the two numbers appearing in this place in $\vectorf$ and $\vectorh$). A corollary of this observation is that if $\vectorf_2$ is a vector sequence of bounded dimension which tends to infinity, then every vector sequence at each (or at each except finitely many) position having dimension smaller or equal to the dimension of $\vectorf_2$ is an asymptotic mix of some $\vectorg_2 \le \vectorf_2$. This corollary gives the right-to-left implication in the lemma.
	
	For the  left-to-right implication, we use Lemma~\ref{lem:no-mix}. Let $d_1$ be such that on an infinite set $X\subseteq\Nat$ of positions $\vectorf_1$ has dimension $d_1$ and $f_2$ has a smaller dimension. By Lemma~\ref{lem:no-mix}, there is a vector  sequence 
	\begin{align*}
		\vectorh : X \to \Nat^{d_1}
	\end{align*}
	of dimension $d_1$ 	which is not an asymptotic mix of any vector sequence of smaller dimension. As we have observed, $\vectorh$ is asymptotically equivalent to some $\vectorg_1 \le \vectorf_1$ (when restricted to positions from $X$), because $\vectorf_1$ tends to infinity on all coordinates. Therefore, $\vectorg_1$ is not an asymptotic mix of any $\vectorg_2 \le \vectorf_2$ on $X$, since such a vector sequence $\vectorg_2$ has strictly smaller dimension. 
	We can arbitrarily extend $\vectorg_1$ to all positions outside of $X$, and still it will not be an asymptotic mix of any $\vectorg_2\leq\vectorf_2$. \end{pr}

\section{Encoding a Minsky machine}
\begin{figure}[htbp]
	\centering
	\includegraphics[scale=0.5]{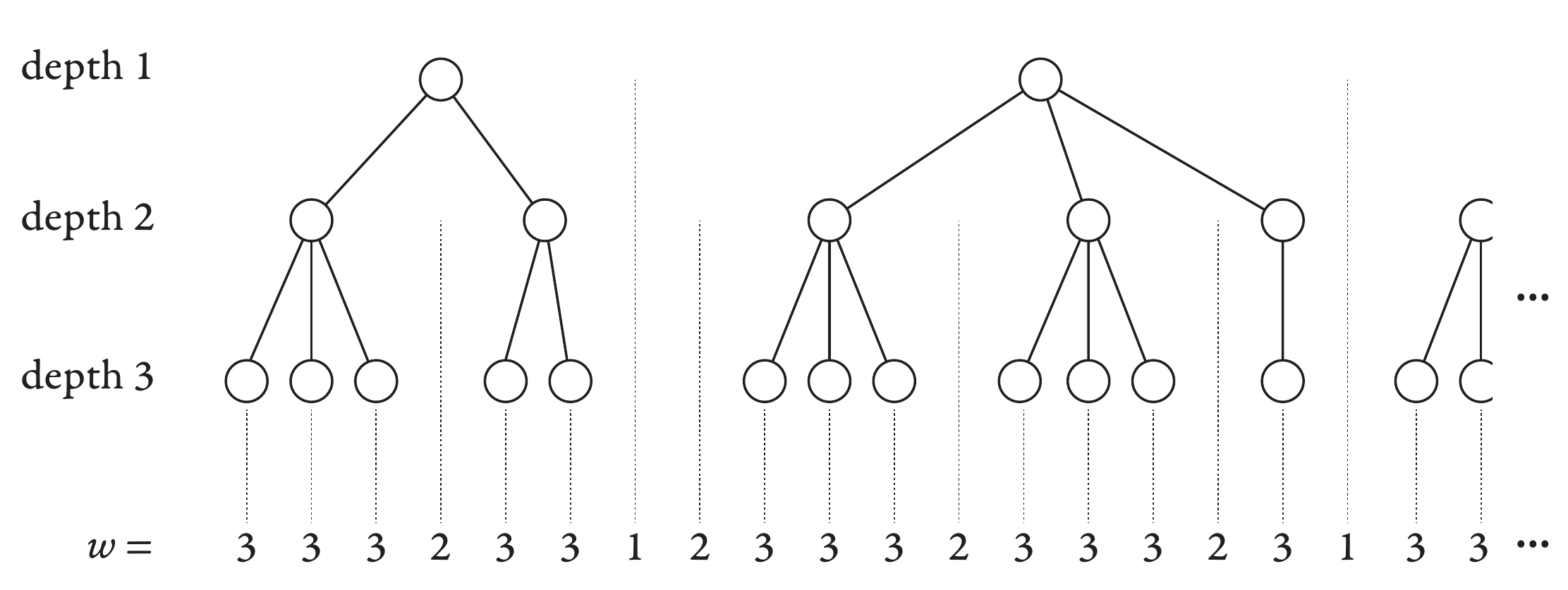}
	\caption{An example of $\tree w$ for $n=3$. Formally speaking, the the leaves of $\tree w$ are positions with label $n$, while the tree structure is defined by the following rule. Two leaves which correspond to positions $x$ and $y$ with label $n$ have a common ancestor at depth $i$ if and only if there is no position between $x$ and $y$ which has label in $\set{1,\ldots,i}$. In particular, if between $x$ and $y$ there is a position with label $1$, then $x$ and $y$ are in different trees of the sequence. Note that the mapping $w \mapsto \tree w$ is not one-to-one, e.g.~in the picture, the first $2$ just after the first $1$ could be removed from $w$ without affecting $\tree w$.
}
	\label{fig:word-to-tree}
\end{figure}

We now use the results on vector sequences from the previous section to prove  undecidability of \msou.  To do this, it will be convenient to view an infinite word as a sequence of finite trees of bounded depth, in the following sense.  Consider a word
\begin{align*}
	w \in \set{1,2,3,\ldots,n}^\omega
\end{align*}
which has infinitely many 1's. We view such a word as an infinite sequence of trees of depth $n$, denoted by $\tree w$, as described in Figure~\ref{fig:word-to-tree}.  % Trees in the sequence are separated by the letter $1$, subtrees of those trees are separated by  letter 2, their subtrees are separated by letter 3, and so on, with letter $n$ representing leaves. This sequence of trees is illustrated below.

The key to the undecidability proof is the following lemma, which says that, in a certain asymptotic sense, degrees can be compared for equality. Here the degree of a tree node is defined to be the number of its children.

\begin{lemma}\label{lem:ruler}
There is an \msou formula, which defines the set of words
\begin{align*}
	w \in \set{1,2,3}^\omega
\end{align*}
which have infinitely many 1's and such that $\tree w$  has the following properties:
\begin{enumerate}[(a)]
	\item\label{it:rulinf} the degree of depth 2 nodes tends to infinity;
	\item\label{it:ruldeg} all but finitely many nodes of depth 1 have the same degree.
\end{enumerate}
\end{lemma}

\begin{pr}
	Condition~(\ref{it:rulinf}) is easily seen to be expressible in \msou. One says that for every set  of depth 2 nodes, their degrees are unbounded. 
	
	Let us focus on condition~(\ref{it:ruldeg}). 	Fix a word $w$ with infinitely many 1's as in the statement of the lemma. For an infinite set $X$ of depth 1 nodes, define 
	\begin{align*}
		\vectorf_X : \Nat \to \Nat^*
	\end{align*}
	to be the vector sequence, where the $i$-th vector is the sequence of degrees of the children of the $i$-th node from $X$. Condition 1 says that if $X$ is the set of all depth 1 nodes, then $\vectorf_X$ tends to infinity, which implies that $\vectorf_X$ also tends to infinity for any other infinite set $X$ of depth 1 nodes. 
	
	Call two sets $X,Y$ of depth 1 nodes \emph{alternating} if every two nodes in $X$ are separated by a node in $Y$, and vice versa. Condition~(\ref{it:ruldeg}) is equivalent to saying that
	\begin{itemize}
		\item depth 1 nodes have bounded degree;
		\item  one cannot find  infinite alternating sets $X,Y$ of depth 1 nodes, such that infinitely often $\vectorf_X$ has strictly bigger dimension than $\vectorf_Y$.
	\end{itemize}
The first condition is clearly expressible in \msou, while the second is expressible in \msou thanks to Lemma~\ref{lem:dimension-char}.
		\end{pr}

\paragraph*{Minsky machines.} To prove undecidability, we reduce emptiness for Minsky machines to deciding \msou. By a Minsky machine we mean a (possibly nondeterministic) device which has a finite state space, and  two counters that can be incremented, decremented,  and tested for zero. It is undecidable if a given Minsky machine has  an accepting run, i.e.~one which begins in a designated initial state with zero on both counters, and ends in a designated final state.

Let $\rho$ be a finite run of a Minsky machine of length $d$.
We say that a vector of natural numbers $(n_1,\ldots,n_{2d})$ \emph{describes} the run $\rho$  if, for $i=1,\ldots,d$,  the numbers $n_{2i-1},n_{2i}$ store the value of the two counters in the $i$-th configuration of $\rho$. Note that this description does not specify fully the run $\rho$, as the state information is missing.
The following lemma contains the reduction of Minsky machine emptiness to  satisfiability of \msou.

\begin{figure}[htbp]
	\centering
		\includegraphics[scale=0.5]{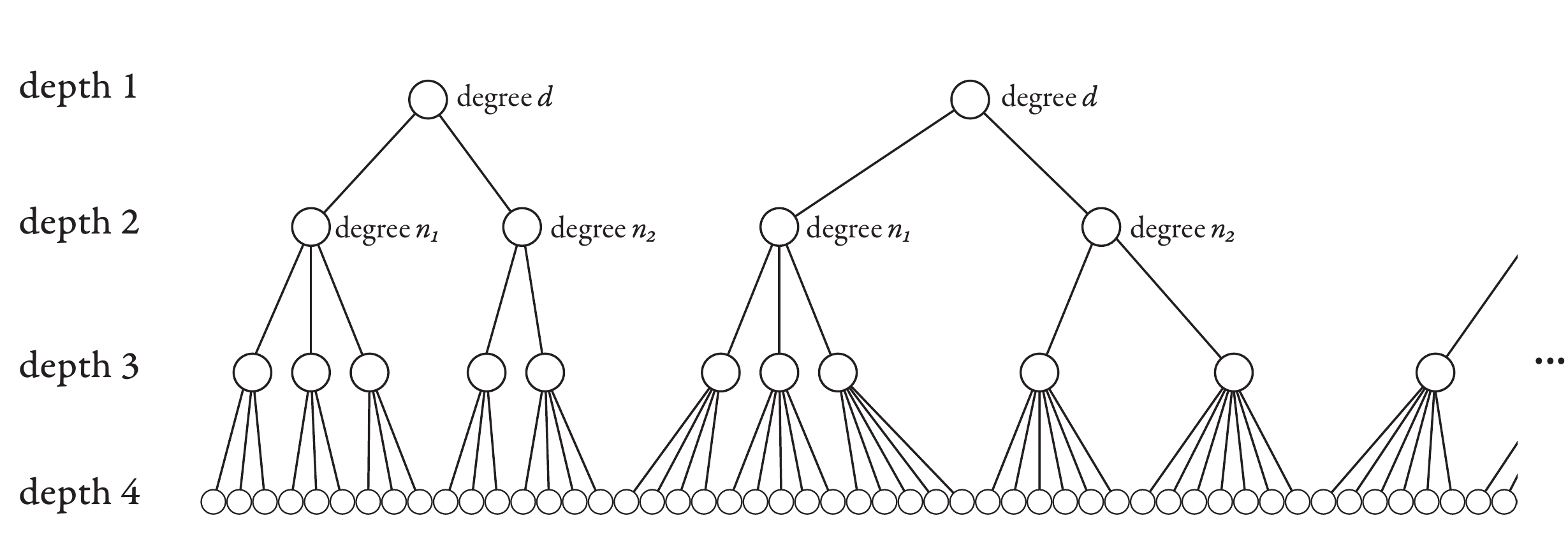}
	\caption{A sequence of trees as in Lemma~\ref{lem:minsky}. Here $d=2$, $n_1=3$, and $n_2=2$. }
	\label{fig:depth4}
\end{figure}
\begin{lemma}\label{lem:minsky}
 For every Minsky machine, one can compute a formula of \msou which defines the set of words
	\begin{align*}
		w \in \set{1,2,3,4}^\omega
	\end{align*}
which have infinitely many 1's and such that $\tree w$ has   the following properties, which are illustrated in Figure~\ref{fig:depth4}:
\begin{enumerate}[(a)]
		\item \label{it:degree-3-infinity} the degree of depth 3 nodes tends to infinity;
	\item \label{it:degree-eventually-constant} all but finitely many depth 1 nodes have the same degree $d$;
	\item \label{it:degree-degree-eventually-constant} for every $i \in \set{1,\ldots,d}$, all but finitely many depth 2 nodes which are an $i$-th child have the same degree, call it $n_i$;
	\item \label{it:two-counter}  $n_1-1,\ldots,n_d-1$ describe some accepting run of the Minsky machine.
\end{enumerate}
\end{lemma}
	\begin{pr}
		Condition~(\ref{it:degree-3-infinity}) is clearly expressible in \msou.  
		
We say that a sequence of trees of depth 3 is \emph{well-formed} if the degree of depth 2 nodes tends to infinity, and that it has \emph{almost constant degree}  if all but finitely many depth 1 nodes have the same degree. Lemma~\ref{lem:ruler}  says that \msou can express the conjunction of being well-formed and having constant degree.  We will use this property to define conditions~(\ref{it:degree-eventually-constant}),~(\ref{it:degree-degree-eventually-constant}) and~(\ref{it:two-counter}).

Define the \emph{flattening} of $\tree w$ to be the sequence of depth 3 trees obtained from $\tree w$ by removing all depth 3 nodes and connecting all depth 4 nodes directly to their depth 2 grandparents. By condition~(\ref{it:degree-3-infinity}), the flattening is well-formed. Since the flattening does not change the degree of depth 1 nodes,  condition~(\ref{it:degree-eventually-constant}) is the same as saying that the flattening has almost constant degree, and therefore can be expressed in \msou thanks to Lemma~\ref{lem:ruler}.

		Define a \emph{depth 2 selector with offset $i$} to be a set of nodes $X$ in the tree $\tree{w}$ which selects  exactly one child for every depth 1 node (and therefore $X$ contains only depth 2 nodes), and  all but finitely many nodes in $X$ are an $i$-th child. A \emph{depth 2 selector}, without $i$ being mentioned,  is a depth 2 selector for some $i$.  Being a depth 2 selector   is equivalent  to saying that one gets a well-formed sequence of almost constant degree if one keeps only nodes from $X_{\leftarrow}$ and their descendants, where $X_{\leftarrow}$ is the set of nodes of depth 2 that have a sibling from $X$ to the right. Therefore, being a depth 2 selector is definable in \msou. Condition~(\ref{it:degree-degree-eventually-constant}) is the same as saying that for every depth 2 selector $X$, if one only keeps the nodes from $X$ and their descendants, then the resulting sequence has almost constant degree, which can be expressed in \msou thanks to Lemma~\ref{lem:ruler}.

We are left with condition~(\ref{it:two-counter}) about Minsky machines.  We say that a depth 2 selector $X$ \emph{represents zero}, if all but finitely many nodes in $X$ have degree one (recall that condition~(\ref{it:two-counter}) uses $n_i-1$ to represent a counter value, because a depth 2 node cannot have degree zero). Representing zero  is definable in first-order logic. 
If $X,Y$ are selectors, we say that \emph{$Y$ increments $X$} if there is some $n$ such that all but finitely many nodes in $X$ have degree $n$, and all but finitely many nodes in $Y$ have degree $n+1$. This is equivalent to saying that if one keeps only nodes from $X \cup Y$ and their descendants, and then removes one subtree of every node from $Y$, then the resulting sequence of depth 3 trees has almost constant degree. Therefore incrementation  is definable in \msou. Using formulas for representing zero and incrementation, it is easy to formalise condition~(\ref{it:two-counter}) in \msou (the formula first guesses the missing state information to fully specify the run $\rho$, and then verifies its consistency with the Minsky machine).
	\end{pr}

In particular, the formula computed in Lemma~\ref{lem:minsky} is satisfiable if and only if the Minsky machine has an accepting run. This yields undecidability of \msou on infinite words, which is the same as our main Theorem~\ref{thm:main}. A corollary of the main theorem is undecidability of the logic {\sc mso}+inf, which is a logic on profinite words defined in~\cite{DBLP:conf/icalp/Torunczyk12}, because decidability of \msou reduces to decidability of {\sc mso}+inf.

\bibliographystyle{alpha}
\bibliography{bib}
\end{document}